\documentclass[10pt]{iopart}
\usepackage{epsfig}
\begin{document}
\title{Breakdown of the KLN Theorem for Charged 
Particles in Condensed Matter}
\author{Y. N. Srivastava\dag , and A. Widom\ddag\footnote[3]
{To whom correspondence should be addressed.}}

\address{\dag\ Department of Physics \& INFN,  University of Perugia, 
Perugia, Italy }

\address{\ddag\ Department of Physics, Northeastern University, 
Boston MA, USA}

\begin{abstract}
The Kinoshita-Lee-Nauenberg (KLN) theorem describes the fact 
that inclusive electromagnetic and weak production processes in the 
vacuum do not contain singularities in the ultra-relativistic limit 
of zero mass. When these production processes occur in condensed matter, 
the KLN theorem fails. One consequence of this failure is that 
precision lifetime determinations of stopped muons will depend on the 
nature of the surrounding material.  
\end{abstract}


\maketitle

\section{Introduction}

The KLN theorem (Kinoshita 1962, Lee and Nauenberg 1964) refers to 
the removable nature of singularities in the production 
probabilities of particles in the limit of zero mass. For example, that 
the photon has zero mass in the vacuum led to the quantum 
electrodynamic ``infrared catastrophe'' wherein more and 
more photons were created at lower and lower frequencies. 
The catastrophe was cured with the realization that photon 
detectors have finite energy resolution. If the resolution 
was kept finite as a formal photon mass went to zero, no 
infinities appeared in the final results.

A generalization may be made for ultra-relativistic charged 
particles wherein the mass appears as a small parameter. For 
example, consider the weak decay
\begin{equation}
\mu^-\to e^-+\bar{\nu}_e+\nu_\mu . 
\end{equation}
 The vacuum {\em radiatively corrected} width (to lowest orders in 
\begin{math} G_F \end{math} and 
\begin{math} \alpha \end{math}) is given by 
(Kinoshita T and Sirlin A 1959, Berman S 1958, Lee T D 1988)
\begin{equation}
\Gamma_\mu =\Gamma^{(0)}_\mu  
\left\{1-{\alpha\over 2\pi }\left(\pi^2-{25\over 4}\right)
-\left({8m^2\over M^2}\right)+\ ... \ \right\},
\end{equation}   
where 
\begin{math}
\Gamma^{(0)}_\mu =
(Mc^2/192\pi^3\hbar)(G_FM^2/\hbar c)^2
\end{math},
\begin{math} M  \end{math} is the muon mass, 
\begin{math} m  \end{math} is electron mass, and 
\begin{math} \alpha =(e^2/\hbar c)  \end{math}. 
Note, in the fictitious zero electron mass limit 
\begin{math} m\to 0 \end{math}, there is no singularity in 
\begin{math} \Gamma_\mu  \end{math}. 
In its general form, the KLN theorem protects the production probability (in 
the vacuum) from zero mass limit singularities. 

However, when a muon decays from rest in condensed matter (as in laboratory 
measurements of \begin{math} \Gamma_\mu  \end{math}) the vacuum KLN protection 
no longer holds true. This becomes evident from the retardation force 
\begin{math}  F  \end{math} in the condensed 
matter felt by an ultra-relativistic electron. 
It is well known theoretically (Berestetskii V B {\it et. al.} 1994) 
and experimentally (Caso C {et. al.} 1998) that  
\begin{equation}
F\approx \left({e^2\omega_p^2\over c^2}\right)
\left\{ln\left({2E^2\over I mc^2}\right)-1+\ ...\right\}, \ \ E>>mc^2 ,
\end{equation}
where \begin{math} I  \end{math} is the log mean {\em ionization potential} 
of the atoms, and the plasma frequency 
\begin{equation} 
\omega_p^2=\left({4\pi ne^2\over m}\right).
\end{equation}
The finite number density \begin{math} n  \end{math} of electrons in the 
material is responsible for decelerating the fast electron in accordance 
with Eq.(3), and by a similar process is responsible for stopping the  
muon in the first place. Clearly, the occurrence of the electron 
mass in the logarithm of Eq.(3) precludes a smooth limit 
\begin{math} m\to 0 \end{math} in the retardation force 
\begin{math} F \end{math} for ultra-relativistic 
particles. The KLN violation is present as long as the density of electrons 
in the condensed matter is finite. This is the first instance for which 
the KLN theorem for decays in condensed matter fails. 

In more detail, the wave function \begin{math} \psi (x)  \end{math} for 
a fast lepton of mass \begin{math} (\hbar \kappa /c) \end{math} traveling 
through a condensed matter medium obeys a non-local Dirac equation 
of the form (Dyson F J 1949, Schwinger J 1951) 
\begin{equation}
\left(-i\gamma^\mu \partial_\mu +\kappa \right)\psi (x)+
\int \Sigma (x-y)\psi (y)d^4 y=0, 
\end{equation}
where the ``self energy'' \begin{math} \Sigma (x-y)  \end{math} depends 
on the continuous medium and is certainly {\em not} well approximated by 
a local potential. The fact that condensed matter induces a non-locality 
in the self energy part is merely another description of the ability of 
condensed matter to stop a high energy particle. To lowest order in 
\begin{math} \alpha \end{math}, the ultra-relativistic lepton has a 
self energy 
\begin{equation}
\Sigma(x-y)=
i\alpha \gamma^\mu S(x-y)\gamma^\nu {\cal D}_{\mu \nu }(x-y)+\ ...,
\end{equation} 
where \begin{math} S(x-y) \end{math} obeys the free lepton Dirac 
equation 
\begin{equation}
\left(-i\gamma^\mu \partial_\mu +\kappa \right)S(x-y)=\delta (x-y),
\end{equation} 
and the photon propagator {\em in the condensed matter medium} 
has the form 
\begin{equation} 
{\cal D}_{\mu \nu }(x)=\int D_{\mu \nu }(Q) e^{iQ\cdot x} 
{d^4 Q\over (2\pi )^4}\ . 
\end{equation} 

In Sec.2, we shall discuss the form of the photon propagator 
\begin{math}  D_{\mu \nu }(Q) \end{math} in continuous media 
described by a dielectric response function 
\begin{math} \varepsilon (\zeta )\end{math} for complex frequency 
\begin{math} \zeta  \end{math}. In Sec.3, it will be shown how 
the retardation force on an ultra-relativistic lepton can be related 
to the condensed matter induced non-local self energy part 
\begin{math} \Sigma (x-y)  \end{math}. This induced self energy 
renders the KLN protection in the limit 
\begin{math} \kappa \to 0 \end{math} invalid. In the concluding Sec.4, 
we explore a notion of how KLN violating condensed matter effects 
can be made manifest in transition rates, 
e.g. \begin{math} \Gamma_\mu  \end{math}.  

\section{Electromagnetic Fields in Continuous Media}

The photon propagator in a material medium will be reviewed. All 
discussions of electromagnetic fields in materials should start with 
the Maxwell's equations 
\begin{equation}
\partial_\mu \ ^*F^{\mu \nu}=0,
\end{equation}
\begin{equation}
\partial_\mu F^{\mu \nu}=-\left({4\pi \over c}\right)J^\nu .
\end{equation}
The current may be broken up into a polarization current plus an 
external current 
\begin{equation}
J^\nu =c\partial_\mu P^{\mu \nu} +J_{ext}^\nu .
\end{equation}
Eq.(11) reads as   
\begin{math} \rho =-div{\bf P}+\rho_{ext}  \end{math} 
and \begin{math} {\bf J} =c\ {\bf curl\  M}+
(\partial {\bf P}/\partial t)+{\bf J}_{ext} \end{math}, in a specific 
Lorentz frame. If we now define 
\begin{equation}
H^{\mu \nu }=F^{\mu \nu }+4\pi P^{\mu \nu},
\end{equation} 
then Eqs.(10)-(12) imply 
\begin{equation}
\partial_\mu H^{\mu \nu}=-\left({4\pi \over c}\right)J_{ext}^\nu .
\end{equation}
Just as \begin{math}  F^{\mu \nu} \end{math} yields   
\begin{math} {\bf E} \end{math} and \begin{math} {\bf B}  \end{math}, 
we have \begin{math}  H^{\mu \nu} \end{math} yielding 
\begin{math} {\bf D}={\bf E}+4\pi {\bf P} \end{math} 
and \begin{math} {\bf H}= {\bf B}-4\pi {\bf M}  \end{math} via Eq.(12).  
Maxwell's Eqs.(9) and (13) determine the electromagnetic fields 
in material media. However the electromagnetic properties of the medium  
must be included as constitutive equations (in linear order) to compute  
the photon propagator. 

The constitutive equations are most easily expressed in the rest frame 
of the condensed matter. However, we here take the option of employing 
a strictly covariant notation. This may be done by considering the 
stress tensor \begin{math} T_{\mu \nu}  \end{math} for the material. 
The stress tensor has four eigenvalues found from  
\begin{math}det||T_{\mu }^{\nu }-P\delta_{\mu }^{\nu }||=0 \end{math}. 
Three of the eigenvalues \begin{math} (P_{I},P_{II},P_{III})  \end{math} 
correspond to possible values of the material pressure in three 
spatial eigenvector directions. The fourth eigenvalue is determined 
by the material mass density 
\begin{math}(P_{IV}/c^2)=-\rho_{mass} \end{math} 
and is described by a time-like eigenvector 
\begin{equation}
N_\mu N^\mu =-1,
\end{equation}
\begin{equation}
T_{\mu }^{\nu}N_\nu =-(\rho_{mass}c^2) N_\mu .
\end{equation}
The dielectric response is a non-local (in space-time) linear 
operator \begin{math} \hat{\varepsilon } \end{math} describing 
\begin{equation}
(N_{\mu }H^{\mu \nu })=\hat{\varepsilon }(N_{\mu }F^{\mu \nu }),
\end{equation} 
i.e. \begin{math} {\bf D}=\hat{\varepsilon }{\bf E} \end{math} in 
the rest frame of the material. We shall further assume that 
\begin{equation}
N_{\mu}\ ^*H^{\mu \nu }=N_{\mu }\ ^*F^{\mu \nu }
\end{equation}
i.e. \begin{math} {\bf H}={\bf B} \end{math} in 
the rest frame of the material.

Thus far our description is {\em strictly gauge invariant} 
since we have not yet introduced vector potentials, and 
manifestly Lorentz covariant. However, in order to discuss 
the photon propagator one must choose a vector potential in a 
particular gauge. One satisfies Eq.(9) by introducing a vector 
potential \begin{math} A^\mu \end{math} via 
\begin{equation}
F_{\mu \nu }=\partial_\mu A_\nu -\partial_\nu A_\mu .
\end{equation}
In the vacuum, it is often convenient to employ the covariant 
(Lorentz) gauge 
\begin{math} \partial_\mu A^\mu_{L}=0 \end{math}.
The close analogue to the covariant gauge is the material 
Lorentz gauge. In the condensed matter rest frame, with 
\begin{math} N^\mu =({\bf 0},1) \end{math} and 
\begin{math} A^\mu =({\bf A},\phi ) \end{math}, we fix the gauge 
according to  
\begin{equation}
{1\over c}\left({\partial (\hat{\varepsilon }\phi ) 
\over \partial t}\right)+div{\bf A}=0.
\end{equation}
In this gauge Eqs.(13) and (16)-(18) read 
\begin{equation}
\left\{{\hat{\varepsilon }\over c^2}
\left({\partial \over \partial t}\right)^2-\nabla ^2\right\}\phi 
=\left(4\pi \over \hat{\varepsilon }\right)\rho_{ext},
\end{equation} 
and 
\begin{equation}
\left\{{\hat{\varepsilon }\over c^2}
\left({\partial \over \partial t}\right)^2-\nabla ^2\right\}{\bf A} 
=\left(4\pi \over c\right){\bf J}_{ext}.
\end{equation}
From Eqs.(20) and (21), we find the material Lorentz gauge 
photon propagator in the space like directions 
\begin{equation}
D_{ij}({\bf Q},\omega )=
\left({
4\pi \delta_{ij}\over 
\big(|{\bf Q}|^2-\varepsilon(|\omega |+i0^+)(\omega /c)^2-i0^+\big)
}\right),
\end{equation}
and in the time-like direction 
\begin{equation}
D_{00}({\bf Q},\omega )=
\left({
4\pi \over \varepsilon(|\omega |+i0^+)
\big(|{\bf Q}|^2-\varepsilon(|\omega |+i0^+)(\omega /c)^2-i0^+\big)
}\right).
\end{equation}
The expressions in Eqs.(22) and (23) for the photon propagator in 
matter in the gauge of Eq.(19) are well known 
(Abrikosov A A {\it et. al.} 1963). These have been fully 
discussed for materials, along with the other standard gauges 
(e.g. radiation or temporal) with regard to the statistical 
physics of radiation in matter (Lifshitz E M and Pitaevskii L P 1984). 
  
\section{The Lepton Self Energy}

In \begin{math}k\end{math}-space, the self energy of Eqs.(6) and (7) has 
the form shown in Fig.1; It is 
\begin{equation}
\Sigma (k)=i\alpha \int\gamma^\mu 
\left({1\over \kappa +\gamma \cdot (k-Q)-i0^+}\right)
\gamma^\nu D_{\mu \nu }(Q){d^4Q\over (2\pi)^4}\ .
\end{equation} 
\begin{figure}
\begin{center}
\epsfig{file=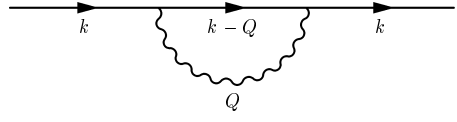,width=6cm}
\end{center}
\caption{The self energy of a lepton is shown  
to lowest order in $\alpha $.}
\end{figure}
If an ultra-relativistic charged lepton having four velocity 
\begin{math} v  \end{math} is moving through the condensed matter 
and is described by a Dirac spinor \begin{math} u \end{math}, 
then the transition rate per unit proper time 
\begin{math} \Gamma_\gamma \end{math} to emit a photon in the medium 
is determined by the imaginary part of the self energy; i.e.  
\begin{math}\Gamma_\gamma 
=-2c\Im m \big(\bar{u} \Sigma (k=\kappa v/c) u \big)\end{math}.
This amounts to the quasi-classical Dirac matrix replacement  
\begin{math}\gamma^\mu \to (v^\mu /c) \end{math}. Thus Eq.(24) 
yields a photon production rate  
\begin{equation}
\Gamma_\gamma 
=2\pi  \alpha \int v^\mu \delta (v\cdot Q)v^\nu 
\Im m \left(D_{\mu \nu }(Q)\right){d^4 Q\over (2\pi )^4} 
\end{equation}
per unit proper time of the ultra-relativistic particle. 

It is of interest to compare our derivation of the photon production 
rate in Eq.(25), via \begin{math} \Sigma (k)  \end{math}, to an 
earlier derivation (Schwinger J 1976) which starts from the non-local 
action for the current, 
\begin{equation}
S=\left({1\over 2c^3}\right)\int \int J^\mu (x) J^\nu (y)
{\cal D}_{\mu \nu}(x-y)d^4x d^4y .
\end{equation} 
If one starts from a ``classical current'' representing the 
ultra-relativistic lepton 
\begin{equation}
J^\mu (x)=ec\int \delta (x-v\tau )v^\mu d\tau , 
\end{equation} 
then radiated photons are created with a Poisson distribution 
\begin{math} p_n=(n^{N_\gamma }/n!)e^{-N\gamma }  \end{math}. 
The mean number of radiated photons is 
\begin{math} N_\gamma  \end{math}. 
The probability of radiating zero photons 
\begin{math} p_0=|e^{iS/\hbar }|^2=e^{-N_\gamma }  \end{math} 
yields 
\begin{math}
N_\gamma =\Im m(2S/\hbar)
\end{math}; i.e. 
\begin{equation}
N_\gamma=\Im m\left({1\over \hbar c^3}\int \int J^\mu (x) J^\nu (y)
{\cal D}_{\mu \nu}(x-y)d^4x d^4y \right).
\end{equation}
With 
\begin{math} 
L^\mu (Q,\tau )=\int_{(-\tau /2)}^{(\tau /2)}
\big(v^\mu e^{iQ\cdot v \tau^\prime }\big)d\tau^\prime , 
\end{math}
Eqs.(8), (27) and (28) yield 
\begin{equation}
N_\gamma =\alpha \Im m\left( \int L^\mu (Q,\tau ) L^\nu (Q,\tau )^* 
D_{\mu \nu }(Q) {d^4 Q\over (2\pi )^4}\right).
\end{equation}
Eq.(29) yields the same number of radiated photons per unit 
proper time 
\begin{equation}
\Gamma_\mu =\lim_{\tau \to \infty}\left({N_\gamma \over \tau }\right) 
\end{equation}
as does Eq.(25), which was based on the self energy part 
\begin{math} \Sigma (k)  \end{math} in Eq.(24).

In the rest frame of the material, employing a laboratory time 
\begin{math} T \end{math}, 
\begin{equation}
\Gamma_\mu^{lab}=\lim_{T\to \infty} 
\left({N_\gamma \over T}\right)
= \Gamma_\mu \sqrt{
1-\left({|{\bf v}|\over c}
\right)^2}\ \ ,
\end{equation}
it follows from Eqs.(22), (23), (25) and (30) that
\begin{equation}
\left({d\Gamma_\mu^{lab}\over d\omega }\right)
=4\pi \alpha {\cal K}(\omega ),\ \ {\rm restricted\ to\ }(\omega >0),
\end{equation}
\begin{equation}
{\cal K}(\omega )=2{\Im m}\int  
\left\{
{ \delta (\omega -{\bf Q\cdot v})
\big(({\bf v}/c)^2-\varepsilon^{-1}(\omega +i0^+)\big)
\over 
\big(|{\bf Q}|^2-\varepsilon(\omega +i0^+)(\omega /c)^2-i0^+\big)}
\right\}
{d^3{\bf Q} \over (2\pi)^3}\ .
\end{equation}
To calculate the retardation force \begin{math} {\bf F} \end{math} 
one may compute the power loss according to 
\begin{equation}
{\bf F\cdot v}=-\int_0^\infty \hbar \omega 
\left({d\Gamma_\mu^{lab}\over d\omega }\right)d\omega =
\end{equation}
\begin{equation}
-\left({2e^2\over \pi v c^2}\right) 
\Im m \int_0^\infty \int_0^{q_{max}}
\left\{
{{\big(
(v/c)^2-\varepsilon^{-1}
\big)\omega q dq d\omega 
\over 
\big(q^2-(\varepsilon -(c/v)^2)(\omega /c)^2-i0^+\big)}}
\right\}\ .
\end{equation}
In Eqs.(34) and (35) we have decomposed 
\begin{math} d^3{\bf Q}=d^2{\bf q}dQ_{||}  \end{math}, 
where \begin{math} Q_{||}=({\bf Q\cdot v}/|{\bf v}|) \end{math}, 
and introduced a high wave number cutoff 
\begin{math} q_{max}  \end{math}.

After playing many ingenious games with logarithms, Eq.(35) in combination 
with the plasma frequency sum rule 
\begin{equation}
\left({2\over \pi }\right)\int_0^\infty 
\omega \Im m\varepsilon (\omega +i0^+)d\omega =\omega_p^2 ,
\end{equation}
yields a high energy retardation force formula in agreement with 
the standard result of Eq.(3), 
\begin{equation}
F\to \left({e^2\omega_p^2\over c^2}\right)
\left\{ln\left({2E^2\over I mc^2}\right)-1+\ ...\right\} \ \ 
{\rm as}\ \ (E/mc^2)\to \infty .
\end{equation}  
The fact that the logarithm in unavoidable in retardation force calculations 
in both theory and laboratory experiments clearly shows the 
failure of the KLN theorem in condensed matter. The theorem is  
(of course) valid and useful for particles moving through the vacuum 
(Sterman G and Weinberg S 1977). But the theorem cannot properly 
be applied to experiments wherein particles move through materials.

\section{Conclusions}

We have shown that the standard condensed matter retardation force 
formulas follow from the self energy part 
\begin{math} \Sigma (k)  \end{math} of a charged lepton. The lowest order 
in \begin{math} \alpha  \end{math} correction is shown in Fig.1. 
Furthermore, the structure of the inclusive transition rate to excite 
the condensed matter, i.e. 
\begin{math} \big(-2c\Im m \Sigma (k)\big)  \end{math}, 
violates the required hypotheses of the KLN theorem. This violation has 
profound implications for ultra-high precision measurements of lepton 
lifetimes.
\begin{figure}
\begin{center}
\epsfig{file=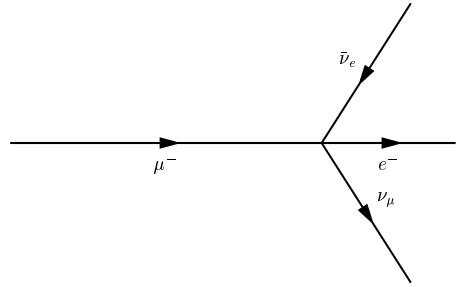,width=6cm}
\end{center}
\caption{The decay of a muon is shown in the bare Fermi theory.}
\end{figure}
\begin{figure}
\begin{center}
\epsfig{file=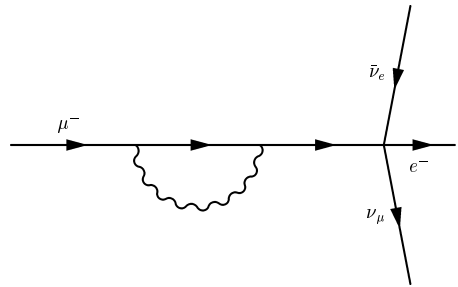,width=6cm}
\end{center}
\caption{The self energy of the muon to lowest order in $\alpha $.}
\end{figure}
\begin{figure}
\begin{center}
\epsfig{file=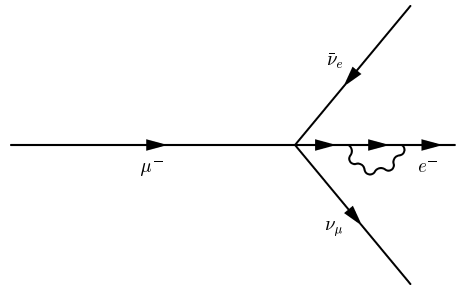,width=6cm}
\end{center}
\caption{The self energy of the electron to lowest order in $\alpha $.}
\end{figure}
\begin{figure}
\begin{center}
\epsfig{file=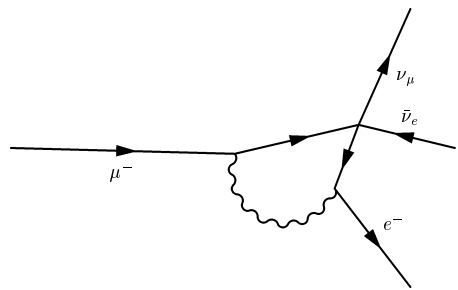,width=6cm}
\end{center}
\caption{The muon-electron vertex correction to lowest 
order in $\alpha $.}
\end{figure}
For example, consider the decays both with and without real photon 
radiation. If the photons are merely virtual, as in Figs.3, 4 and 5, 
then the reaction is in reality   
\begin{equation}
\mu^-\to e^- +\nu_\mu +\bar{\nu }_e .
\end{equation}  
From the interference of the superposition of amplitudes shown in 
Figs.3, 4 and 5, with the amplitude shown in Fig.2, one finds a 
{\em decrease} in the total muon decay rate. On the other hand 
there is a somewhat smaller increase in the total muon decay 
rate due the emission of real photons as shown in Figs.6 and 7. 
The real photon emission reaction is  
\begin{equation}
\mu^-\to e^- +\nu_\mu +\bar{\nu }_e +\gamma .
\end{equation} 
\begin{figure}
\begin{center}
\epsfig{file=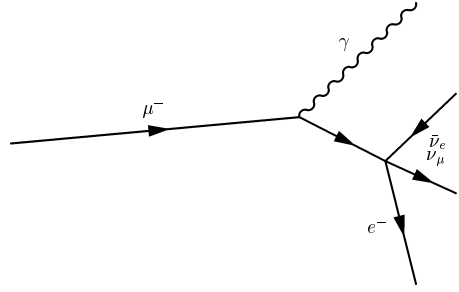,width=6cm}
\end{center}
\caption{The muon emits a real photon.}
\end{figure}
\begin{figure}
\begin{center}
\epsfig{file=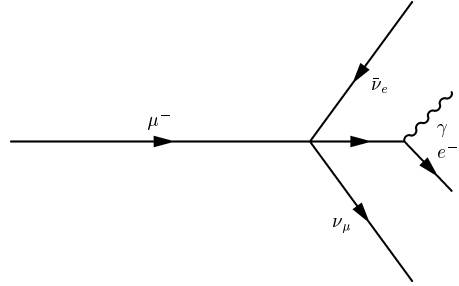,width=6cm}
\end{center}
\caption{The electron emits a real photon.}
\end{figure}
Both Eqs.(38) and (39) should include radiative corrections of order 
\begin{math} \alpha \end{math}. The total effect to order 
\begin{math} \alpha \end{math} is to {\em decrease} the total decay 
rate of the muon.

In order to properly include initial and final state interactions, 
the self energy parts \begin{math} \Sigma (k) \end{math} of both the muon 
and the electron, i.e. the diagrams both in Figs.3 and 4 must be 
calculated {\em including condensed matter effects} as fully discussed 
in this work. It makes {\em no sense} to ignore the simple experimental 
fact that charged particles may be stopped in condensed matter via the 
induced lepton self energy parts \begin{math} \Sigma (k) \end{math}.

Once this simple experimental fact is realized, it becomes evident 
(for reasons of gauge invariance) that all photons, virtual or real, 
must be renormalized by the condensed matter. Since all photons 
propagate as Eqs.(20) and (21), condensed matter effects must be included 
in the final state interactions of the photon. Thus, the condensed 
matter corrections up to order \begin{math} \alpha \end{math}  
shown become somewhat larger than the vacuum terms of order 
\begin{math} \alpha^2 \end{math}. 

The breakdown of the KLN theorem in condensed matter dictates 
that internal electromagnetic fields must be seriously considered 
in precision computations of experimental muon lifetimes. 
The condensed matter electric fields are sufficiently large to 
actually stop an energetic lepton. This would be most difficult 
to do in a vacuum. The requirement of considering condensed matter 
effects in precision determinations of the muon lifetime is most 
easily proved (or disproved) by experimental studies in which the 
muons are stopped in different sorts of materials.

\section*{References}
\begin{harvard}
\item[] Kinoshita T 1962 {\it J. Math Phys. }{\bf 3} p-650
\item[] Lee T D and Nauenberg M 1964 {\it Phys. Rev. }{\bf 133} 
p-B1459 
\item[] Kinoshita T and Sirlin A 1959 {\it Phys. Rev. }{\bf 113} 
p-1652
\item[] Berman S 1958 {\it Phys. Rev. }{\bf 112} p-267  
\item[] Lee T D 1988, ``Particle Physics and Introduction to 
Field Theory'', Harwood Academic Publishers (New York)
\item[] Berestetskii V B, Lifshitz E M and Pitaevskii L P 1994, 
``Quantum Electrodynamics'', Pergamon Press (Oxford), Eq.(82.26) 
p-335
\item[] Caso C 1998, {\it et. al.}, Review of Particle Physics, 
{\it Euro. Phys. J. }{\bf C3} p-1
\item[] Dyson F J 1949 {\it Phys. Rev. }{\bf 75} p-1736
\item[] Schwinger J 1951 {\it Proc. Nat. Acad. Sciences }{\bf 37} 
p-452 
\item[] Abrikosov A A, Gorkov L P and Dzyaloshinski I E 1963, 
``Methods of Quantum Field Theory in Statistical Physics'' 
Dover Publications Inc. (New York) ch-6
\item[] Lifshitz E M and Pitaevskii L P 1984, 
``Statistical Physics Part 2'', Pergamon Press (Oxford), sec-76.
\item[] Schwinger J 1976 in ``Gauge Theories and Modern Field 
Theory'', ed. Arnowitt R and Nath P, MIT Press (Cambridge) p-337 
\item[] Sterman G and Weinberg S 1977, {\it Phys. Rev. Lett. }
{\bf 39} p-1436

\end{harvard}

\end{document}